\documentclass{osa-article}

\journal{osac}
\articletype{Research Article}
\begin{document}

\title{X-ray verification of sol-gel resist shrinkage in substrate-conformal imprint lithography for a replicated blazed reflection grating}

\author{Jake A.\ McCoy,\authormark{1,*} Marc A.\ Verschuuren,\authormark{2} Drew M.\ Miles,\authormark{1} and Randall L.\ McEntaffer\authormark{1}}

\address{\authormark{1}Department of Astronomy \& Astrophysics, The Pennsylvania State University, 525 Davey Laboratory, University Park, PA 16802, USA\\
\authormark{2}Philips SCIL Nanoimprint Solutions, De Lismortel 31, 5612 AR, Eindhoven, The Netherlands}

\email{\authormark{*}jam1117@psu.edu} %% email address is required



\begin{abstract}
 Surface-relief gratings fabricated through nanoimprint lithography (NIL) are prone to topographic distortion induced by resist shrinkage. 
 Characterizing the impact of this effect on blazed diffraction efficiency is particularly important for applications in astrophysical spectroscopy at soft x-ray wavelengths ($\lambda \approx 0.5 - 5$~nm) that call for the mass-production of large-area grating replicas with sub-micron, sawtooth surface-relief profiles. 
 A variant of NIL that lends itself well for this task is substrate-conformal imprint lithography (SCIL), which uses a flexible, composite stamp formed from a rigid master template to imprint nanoscale features in an inorganic resist that cures thermodynamically through a silica sol-gel process. 
 While SCIL enables the production of several hundred imprints before stamp degradation and avoids many of the detriments associated with large-area imprinting in NIL, the sol-gel resist suffers shrinkage dependent on the post-imprint cure temperature.  
 Through atomic force microscopy and diffraction-efficiency testing at beamline 6.3.2 of the Advanced Light Source, the impact of this effect on blaze response is constrained for a $\sim$160-nm-period grating replica cured at 90$^{\circ}$C. 
 Results demonstrate a $\sim$2$^{\circ}$ reduction in blaze angle relative to the master grating, which was fabricated by anisotropic wet etching in $\langle 311 \rangle$-oriented silicon to yield a facet angle close to 30$^{\circ}$. 
\end{abstract}

%%%%%%%%%%%%%%%%%%%%%%%%%%  body  %%%%%%%%%%%%%%%%%%%%%%%%%%
\section{Introduction}\label{sec:intro}
%%%%%%%%%%%%%%%%%%%%%%%%%%%%%%%%%%%%%%%%%--------------------------------------------------
Instrument development for astrophysical spectroscopy at soft x-ray wavelengths ($\lambda \approx 0.5 - 5$~nm) represents an active area of research that utilizes blazed gratings with sub-micron periodicities, which are often replicated from a master grating template featuring a custom groove layout \cite{McEntaffer13,Miles18,McEntaffer19}. 
Starting with a master grating fabricated by anisotropic wet etching in mono-crystalline silicon and surface-treated for anti-stiction, a sawtooth surface-relief mold that enables high diffraction efficiency in the soft x-ray can be patterned in ultraviolet (UV)-curable, organic resist via UV-nanoimprint lithography (UV-NIL) \cite{Haisma96,Chang03,Miles18}. 
This has been demonstrated by Miles,~et~al.~\cite{Miles18} through beamline diffraction-efficiency testing of a gold-coated, UV-NIL replica with a periodicity of $\sim$160~nm, which was imprinted from a stamp wet-etched in $\langle 311 \rangle$-oriented silicon to yield a nominal blaze angle of 29.5$^{\circ}$ over a 72~cm$^2$ variable-line-space groove layout. 
\begin{figure*}
 \centering
 \includegraphics[scale=0.405]{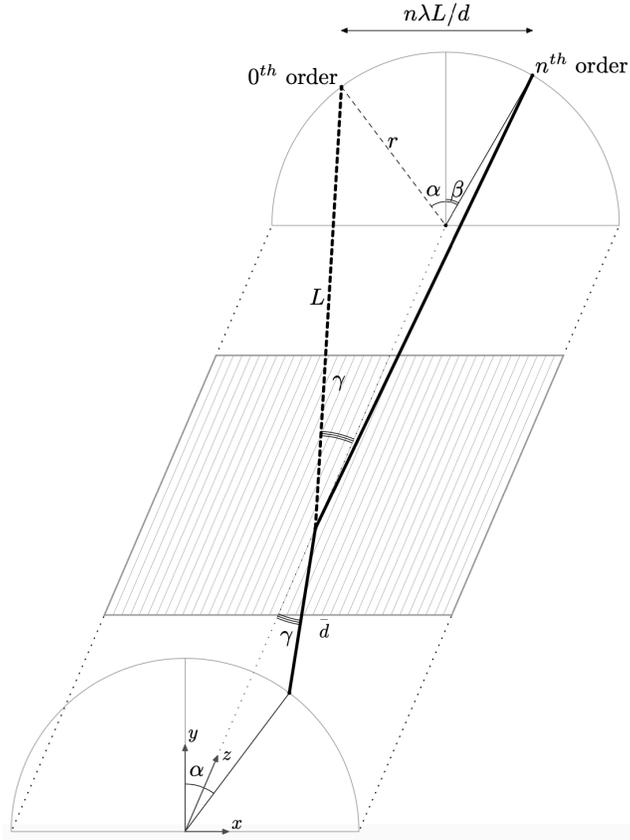}
 \caption{Geometry for a reflection grating producing a conical diffraction pattern \cite{Miles18,McCoy20}. In an extreme off-plane mount, the incoming radiation is nearly parallel to the groove direction with $\gamma \lessapprox 2^{\circ}$ while $\alpha$ is free to match the blaze angle, $\delta$, in a Littrow configuration with $\alpha = \beta = \delta$. At a distance $L$ away from the point of incidence on the grating, the arc radius is $r = L \sin \left( \gamma \right)$ and diffracted orders are each separated by a distance $\lambda L / d$ along the dispersion direction, where $d$ is the groove spacing.}\label{fig:conical_reflection} 
 \end{figure*}
These results show that crystallographic etching coupled with UV-NIL processing is capable of producing large-area, blazed gratings that perform with high diffraction efficiency in an extreme off-plane mount. 
As illustrated in Fig.~\ref{fig:conical_reflection}, the incoming radiation in this geometry is nearly parallel to the groove direction so that propagating orders are confined to the surface of a cone as described by 
\begin{equation}\label{eq:off-plane_incidence_orders}
 \sin \left( \alpha \right) + \sin \left( \beta \right) = \frac{n \lambda}{d \sin \left( \gamma \right)} \; \; \text{for} \; \; n = 0, \pm 1, \pm 2, \pm 3 ... 
 \end{equation}
where $d$ is the groove spacing, $\gamma \lessapprox 2^{\circ}$ is the half-opening angle of the cone, $\alpha$ is the azimuthal incidence angle and $\beta$ is the azimuthal diffracted angle of the $n^{\text{th}}$ diffracted order \cite{Neviere78}. 

While UV-NIL has been proven to be a suitable technology for replicating surface-relief molds for x-ray reflection gratings \cite{McEntaffer13,Miles18}, there are aspects of this process that lead to practical difficulties for realizing a state-of-the-art grating spectrometer with mass-produced reflection gratings. 
First, the rigidity of a thick silicon stamp requires a relatively high applied pressure for imprints of substantial area to achieve conformal contact between the stamp and the resist-coated blank substrate so that air pockets that give rise to unpatterned areas can be avoided \cite{Schift10}. 
High-pressure imprinting conditions can also lead to imperfections that arise from particulate contaminants, and potentially, damage to the stamp surface. 
Additionally, the pattern fidelity of a rigid stamp is gradually degraded as it makes repeated imprints such that in the case of the UV-NIL process described by Miles,~et~al.~\cite{Miles18}, a single stamp typically can produce tens of quality grating replicas \cite{Schift10}. 
As a result, the implementation of UV-NIL becomes impractical for future astronomical instruments such as \emph{The Rockets for Extended-source X-ray Spectroscopy} \cite{Miles19b} and \emph{The Off-plane Grating Rocket Experiment} \cite{Tutt18} that each require hundreds of replicated gratings and additionally, the \emph{X-ray Grating Spectrometer} for the \emph{Lynx X-ray Observatory} mission concept, which calls for the production of thousands of replicated gratings \cite{McEntaffer19}. 

An alternative NIL technique for the mass production of x-ray reflection gratings is substrate-conformal imprint lithography (SCIL) \cite{Verschuuren17,Verschuuren18,Verschuuren19}. 
Unlike standard NIL that uses a rigid stamp for direct imprinting, SCIL centers on the use of a low-cost, flexible stamp molded from a rigid master template. 
With stamp features carried in a modified form of polydimethylsiloxane (PDMS) that has an increased Young's modulus relative to that of standard PDMS, SCIL offers a way for nanoscale patterns to be imprinted in resist over large areas using a stamp that conforms locally to particulate contaminants and globally to any slight bow of the replica substrate, while avoiding damage to the master template by eliminating the need for an applied high pressure. 
Additionally, wave-like sequential imprinting, which is made possible by specialized pneumatic tooling coupled with the flexibility of the stamp, serves to eliminate large trapped air pockets \cite{Verschuuren17,Verschuuren19}. 
Packaged equipment that automates spin-coating and this pneumatic-based SCIL wafer-scale imprint method for high-volume replication has been developed by Philips SCIL Nanoimprint Solutions \cite{scil_nano}. 
This production platform, known as \textsc{AutoSCIL}, was first applied to x-ray reflection grating technology for the grating spectrometer on board the \emph{Water Recovery X-ray Rocket} \cite{Miles19,Verschuuren18}, which utilized 26 nickel-coated replicas of a 110~cm$^2$ master grating fabricated through crystallographic etching in a manner similar to the processing described by Miles,~et~al.~\cite{Miles18}. 

Although SCIL stamps are compatible with many UV-curable, organic resists similar to those used for UV-NIL \cite{Ji10,Schift10}, high-volume production that relies on long stamp lifetime is best suited for use with a brand of inorganic resist that cures through a thermodynamically-driven, silica sol-gel process \cite{Verschuuren17}.
Synthesized by Philips SCIL Nanoimprint Solutions and known commercially as \textsc{NanoGlass}, this resist is stored as a $-20^{\circ}$C sol containing silicon precursors tetramethylorthosilicate (TMOS) and methyltrimethoxysilane (MTMS) suspended in a mixture of water and alcohols \cite{Verschuuren19}. 
When a SCIL stamp is applied to a wafer freshly spin-coated with a film of resist, its features are filled through capillary action while the precursors react to form a gel, and ultimately a solid silica-like network, along with alcohols and water left as reaction products. 
This sol-gel process carries out over the course of 15~minutes at room temperature (or, a few minutes at $\sim$50$^{\circ}$C) while reaction products and trapped air diffuse into the stamp, leaving solidified resist molded to the inverse of the stamp topography after stamp separation. 
The imprinted resist initially has $\sim$70\% the density of fused silica due to the presence of nanoscale pores and methyl groups bound to silicon that arise from the organically-modified MTMS precursor. 
However, the material can be densified for stability through a 15-minute bake at a temperature $T \gtrapprox 50^{\circ}$C to induce further cross-linking in the silica network, where $T \gtrapprox 450^{\circ}$C breaks the silicon-carbon bonds while inducing a moderate level of shrinkage and $T \gtrapprox 850^{\circ}$C gives rise to the density of maximally cross-linked fused silica \cite{Verschuuren19}. 

Using the \textsc{AutoSCIL} production platform, a single stamp is capable of producing $\gtrapprox$700 imprints in sol-gel resist at a rate of 60, 150-mm-diameter wafers per hour, without pattern degradation \cite{Verschuuren17,Verschuuren18,Verschuuren19}. 
While this makes SCIL an attractive method for mass producing x-ray reflection gratings, the thermally-induced densification of the silica sol-gel network causes resist shrinkage similar in effect to the UV-curing of organic resists in UV-NIL \cite{Schift10,Shibata10,Horiba_2012}.  
It has been previously reported that a $T \approx 200^{\circ}$C treatment of sol-gel resist leads to $\sim$15\% volumetric shrinkage in imprinted laminar gratings while temperatures in excess of $1000^{\circ}$C result in a maximal, $\sim$30\% shrinkage \cite{Verschuuren19}.   
Based on these results, it is hypothesized that a low-temperature treatment should lead to $\sim$10\% volumetric shrinkage in the resist, which is comparable to typical levels of resist shrinkage in UV-NIL \cite{Schift10}. 
To probe the impact that resist shrinkage in SCIL has on blaze angle in an x-ray reflection grating, this paper presents beamline diffraction-efficiency measurements of a gold-coated imprinted that was cured at a temperature of $T \approx 90^{\circ}$C and compares them to theoretical models for diffraction efficiency that characterize the expected centroids for peak orders, as well as measurements of the corresponding silicon master grating in a similar configuration. 
These results corroborate atomic force microscopy (AFM) measurements of the tested gratings that, together, serve as experimental evidence for resist shrinkage affecting the blaze response of an x-ray reflection grating through a reduction in facet angle. 

This paper is organized as follows: 
section~\ref{sec:grating_fab} describes the fabrication of the gratings used for this study while section~\ref{sec:beamline_testing} presents their diffraction-efficiency measurements, which were gathered at beamline 6.3.2 of the Advanced Light Source (ALS) synchrotron facility at Lawrence Berkeley National Laboratory (LBNL) \cite{als632,Gullikson01,Underwood96}. 
Section~\ref{sec:discussion} then analyses these results and compares them to AFM measurements in order to demonstrate a non-negligible blaze angle reduction that is expected to occur in the replica based on an approximate model for resist shrinkage. 
Conclusions and a summary of this work are then provided in section~\ref{sec:conclusion}. 
The SCIL processing described in this paper was performed by Philips SCIL Nanoimprint Solutions using a master grating template fabricated at the Nanofabrication Laboratory of the Pennsylvania State University (PSU) Materials Research Institute \cite{psu_mri}. 
All field-emission scanning electron microscopy (FESEM) was carried out with a \textsc{Zeiss Leo 1530} system at the PSU Nanofabrication Laboratory while all AFM was carried out using a \textsc{Bruker Icon} instrument equipped with a \textsc{SCANASYST-AIR} tip and \textsc{PeakForce Tapping}$^{\text{TM}}$ mode at the PSU Materials Characterization Laboratory. 

\section{Grating Fabrication by SCIL}\label{sec:grating_fab}
%%%%%%%%%%%%%%%%%%%%%%%%%%%%%%%%%%%%%%%%%-------------------------------------------------- 
The master grating template chosen for this study was originally used as a direct stamp for the UV-NIL processing described by Miles,~et~al.~\cite{Miles18}. 
This 75~mm by 96~mm (72~cm$^2$) grating was fabricated through a multi-step process centering on anisotropic wet etching in a $\langle 311 \rangle$-oriented, 500-$\mu$m-thick, 150-mm-diameter silicon wafer using potassium hydroxide (KOH). 
As described by Miles,~et~al.~\cite{Miles18}, the groove layout was patterned as a variable-line-space profile using electron-beam lithography with the groove spacing, $d$, ranging nominally from 158.25~nm to 160~nm along the groove direction, which is aligned with the $\langle 110 \rangle$ direction in the $\{ 311 \}$ plane of the wafer surface. 
This layout was then transferred by reactive ion etch into a thin film of stoichiometric silicon nitride (Si$_3$N$_4$) formed by low-pressure chemical vapor deposition before the native silicon dioxide (SiO$_2$) on the exposed surface of the silicon wafer was removed with a buffered oxide etch. 
\begin{figure}
 \centering
 \includegraphics[scale=0.42]{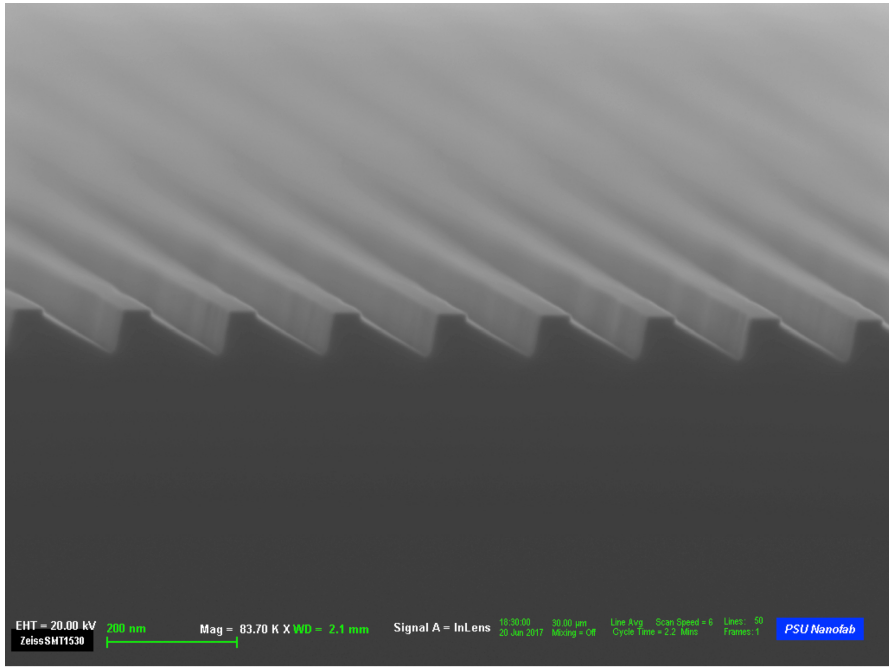}
 \caption{Cross-section FESEM image of the silicon master used for SCIL stamp construction, which was originally used as a direct stamp for UV-NIL \cite{Miles18}.}\label{fig:master_SEM} 
 \end{figure}
Next, a timed, room-temperature KOH etch was carried out to generate an asymmetric, sawtooth-like structure defined by exposed $\{ 111 \}$ planes that form an angle $\theta \equiv \arccos \left( 1/3 \right) \approx 70.5^{\circ}$ at the bottom of each groove, as well as $\sim$30-nm-wide flat-tops that exist beneath the Si$_3$N$_4$ hard mask. 
Due to the $\langle 311 \rangle$ surface orientation of the silicon wafer, the exposed $\{ 111 \}$ planes define nominal facet angles of $\delta = 29.5^{\circ}$ and $180^{\circ} - \theta - \delta \approx 80^{\circ}$. 
A cross-section image of the grating following the removal of Si$_3$N$_4$ using hydrofluoric acid is shown under FESEM in Fig.~\ref{fig:master_SEM}. 

\begin{figure}
 \centering
 \includegraphics[scale=0.4]{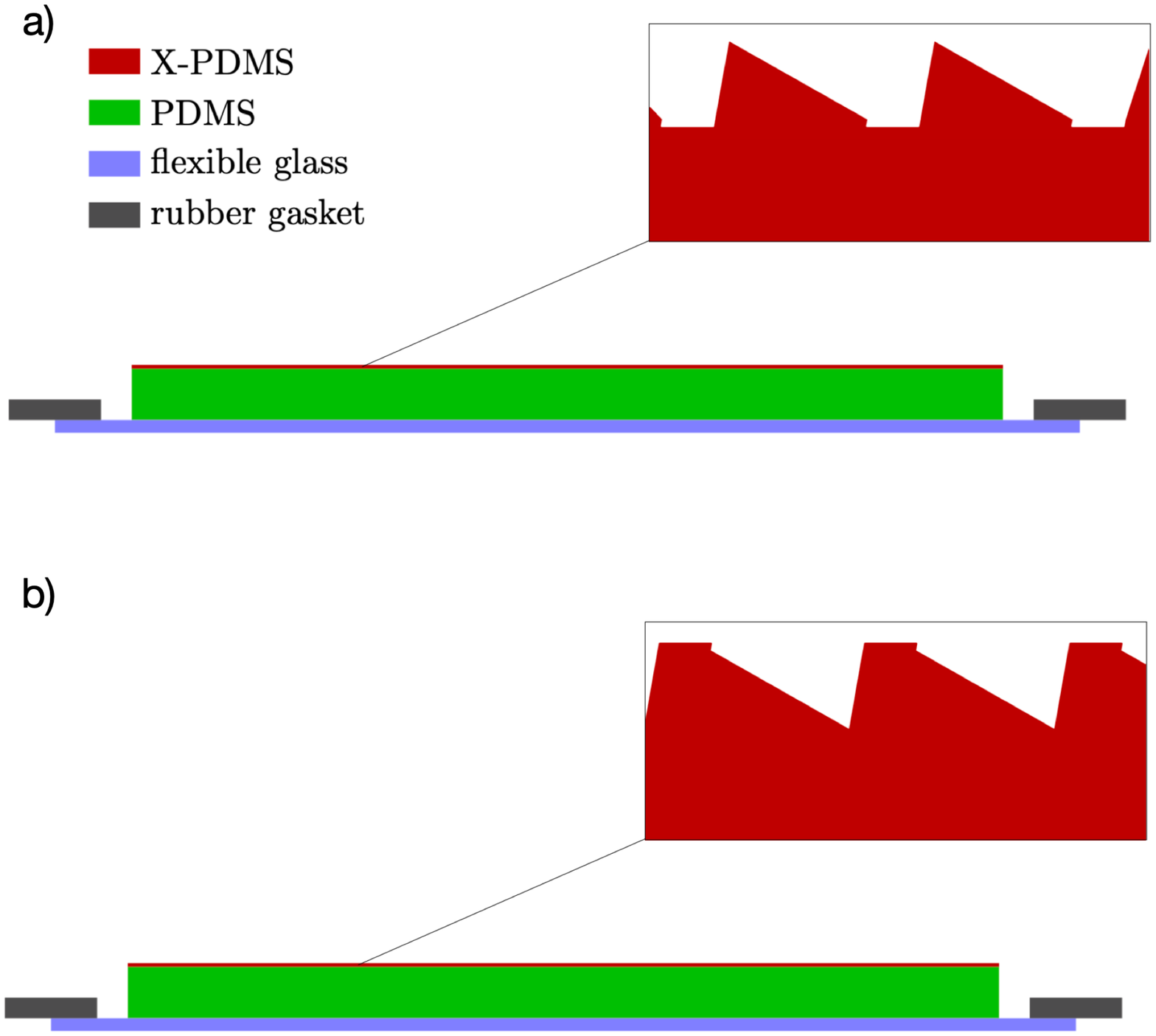}
 \caption{Schematic for SCIL composite stamps of two varieties: a) an initial stamp featuring an inverted topography molded directly from the silicon master shown in Fig.~\ref{fig:master_SEM} and b) a secondary stamp  featuring a topography similar to the silicon master, which was molded using the first stamp as a master template. In either case, grating grooves are carried in a layer of X-PDMS tens of microns thick that sits on a 200-mm-diameter, flexible glass sheet buffered by a $\gtrapprox$0.5-mm-thick layer of soft PDMS. A rubber gasket can be attached for use with the pneumatic-based SCIL wafer-scale imprint method. This illustration neglects slight rounding that can occur in sharp corners under the influence of surface tension in X-PDMS.}\label{fig:scil_stamps}
 \end{figure}

Prior to constructing the composite stamp used for imprint production, the silicon master was cleaned in a heated bath of \textsc{Nano-Strip}$^{\text{TM}}$ (\textsc{VWR Int.}), which consists primarily of sulfuric acid, and then by oxygen plasma before being surface treated for anti-stiction with a self-assembled monolayer of 1,1,2,2H-perfluorodecyltrichlorosilane (FDTS) \cite{Zhuang07} achieved through a 50$^{\circ}$C molecular vapor deposition (MVD) process. 
As described by Verschuuren,~et~al.~\cite{Verschuuren17} and illustrated schematically in Fig.~\ref{fig:scil_stamps}(a), a standard SCIL stamp consists primarily of two components that are supported by a flexible sheet of glass with a thickness of about 200~$\mu$m: a $\sim$50-$\mu$m-thick layer of modified PDMS that carries the inverse topography of the silicon master, and an underlying, $\gtrapprox$0.5-mm-thick layer of standard, soft PDMS that attaches to the glass sheet by application of an adhesion promoter. 
A rubber gasket can then be glued to the outer perimeter of the square glass sheet for use with the pneumatic-based SCIL wafer-scale imprint method to produce imprints with topographies that resemble that of the silicon master. 
However, in an effort to produce imprints that emulate the UV-NIL replica described by Miles,~et~al.~\cite{Miles18}, which was fabricated using the silicon master as a direct stamp, this process was modified to realize a stamp with an inverted topography, as in Fig.~\ref{fig:scil_stamps}(b), so as to allow the production of imprints with sharp apexes and flat portions at the bottom of each groove \cite{Verschuuren18}. 

The variety of modified PDMS used for this study was X-PDMS version~3, (Philips SCIL Nanoimprint Solutions), which was dispensed over the surface of the MVD-treated silicon master and then solidified through two rounds of spin-coating and baking steps using primary and accompanying components of the material. 
First, after the silicon master was cleaned again using deionized water and IPA, $\sim$3~g of the primary component was dispensed over the wafer through a short, 2~krpm spin-coat process using a low spin acceleration, leaving a layer tens of microns thick. 
This was followed immediately by a 50$^{\circ}$C hotplate bake for 3~minutes and a room-temperature cool-down of 10~minutes to leave the material in a tacky state. 
Next, $\sim$3~g of the accompanying component was spin-coated over this layer in a similar way before the wafer was baked by 70$^{\circ}$C hotplate for 10~minutes to form an intermediate layer also tens of microns thick. 
The doubly-coated silicon master was then oven-baked at 75$^{\circ}$C for 20~hours to form a $\sim$50-$\mu$m-thick layer of cured X-PDMS with a Young's modulus on the order of several tens of megapascals. 
In principle, this level of stiffness is sufficient for the stamp to carry grating grooves with $d \lessapprox 160$~nm without pattern distortion or feature collapse \cite{Verschuuren17,Verschuuren19}. 

Using the SCIL Stamp Making Tool (SMT) built by Philips SCIL Nanoimprint Solutions, the initial, non-inverted stamp was formed by curing soft, \textsc{Sylgard 184} PDMS (\textsc{Dow, Inc.}) between the X-PDMS layer and a 200-$\mu$m-thick sheet of D~263 glass (\textsc{Schott AG}), cut into a 200-mm-diameter circle. 
Consisting primarily of two, opposite-facing vacuum chucks heated to 50$^{\circ}$C with surfaces flat to $\lessapprox$10~$\mu$m peak-to-valley, this tool was used to spread $\sim$12~g of degassed PDMS evenly over the surface of the X-PDMS-coated silicon master. 
With the silicon master secured to the bottom chuck and the glass sheet secured to the top chuck, the two components were carefully brought into contact to spread the PDMS to a uniform thickness of $\gtrapprox 0.5$~mm using micrometer spindles, while ensuring that the two surfaces were parallel to within 20~$\mu$m.  
These materials were baked in this configuration at 50$^{\circ}$C until the PDMS was cured so that the stamp could be carefully separated from the silicon master. 
Using this initial stamp as a master template, the secondary, inverted stamp was constructed on a square sheet of glass through steps identical to those outlined above. 
This processing was enabled by the initial stamp being constructed on a round sheet of glass, which allowed it to be spin-coated with X-PDMS and subsequently cured using the same processing steps outlined above for the silicon master. 

Several blazed grating molds were imprinted by hand into $\sim$100-nm-thick films of \textsc{NanoGlass T1100} sol-gel resist spin-coated on 1-mm-thick, 150-mm-diameter silicon wafers using the inverted SCIL stamp just described. % in section~\ref{sec:stamp_construction}. 
\begin{figure*}
 \centering
 \includegraphics[scale=0.42]{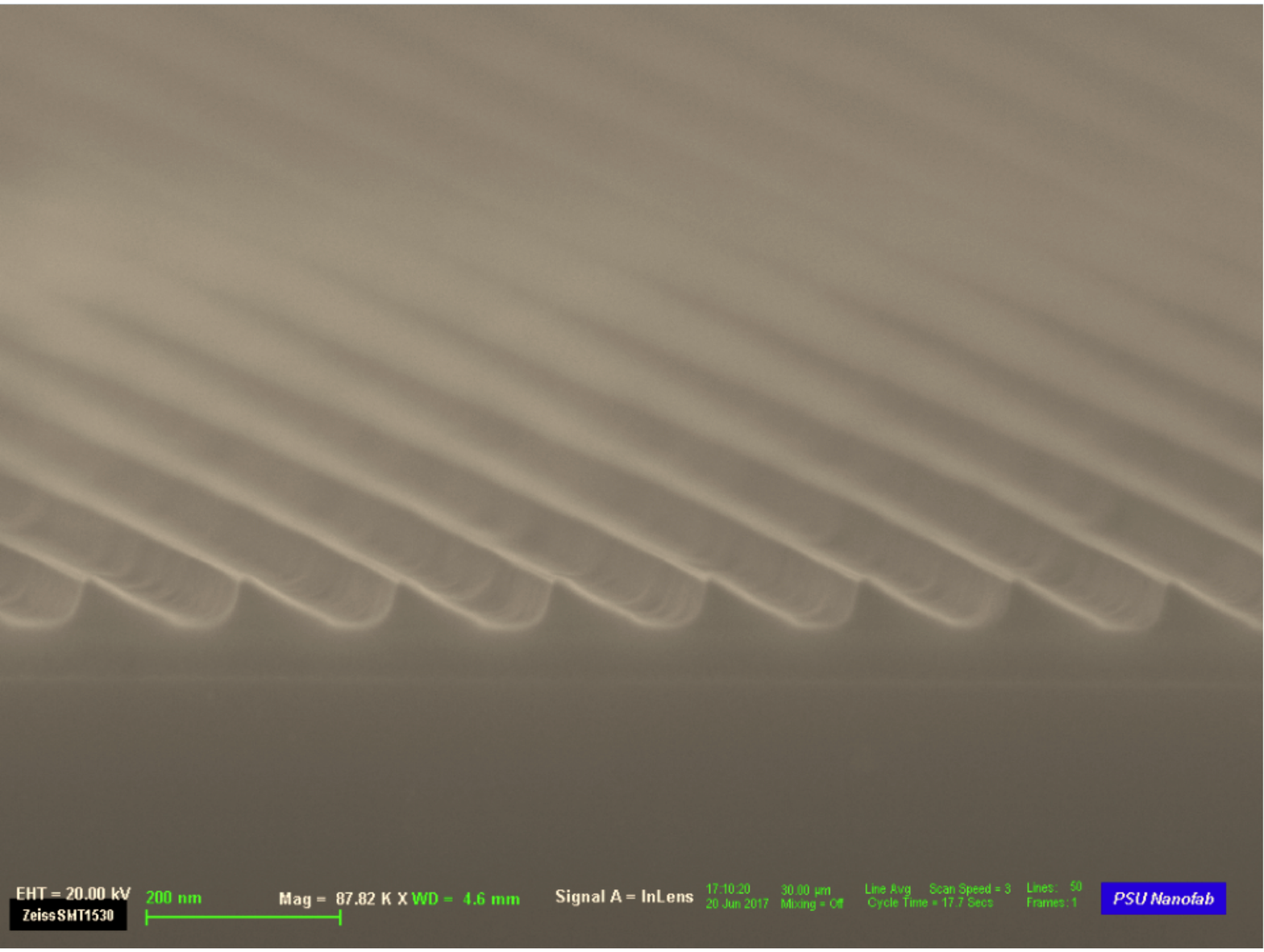}
 \caption{Cross-section FESEM image of a grating imprint with a groove spacing of $d \lessapprox 160$~nm in $\sim$100-nm-thick sol-gel resist coated on a silicon wafer.}\label{fig:SCIL_FESEM} 
 \end{figure*}
Although the pneumatic-based SCIL wafer-scale imprint method is best equipped for minimizing pattern distortion over 150-mm-diameter wafers, imprinting by hand is sufficient for producing a small number of grating molds suitable for the diffraction-efficiency testing described in section~\ref{sec:beamline_testing}, which depends primarily on the groove facet shape over a local area defined by the projected size of the monochromatic beam at the ALS. 
With imprinting taking place at room temperature, 15~minutes of stamp-resist contact was allotted for the sol-gel process to carry out in each imprint. 
Each wafer was baked by hotplate to 90$^{\circ}$C for 15~minutes following stamp separation to densify the imprinted material to a small degree, thereby inducing resist shrinkage. 
An FESEM cross-section of a replica produced in this way is shown in Fig.~\ref{fig:SCIL_FESEM}, where grating grooves are seen imprinted over a residual layer of resist a few tens of nanometers thick. 

\section{Beamline Experiments}\label{sec:beamline_testing}
%%%%%%%%%%%%%%%%%%%%%%%%%%%%%%%%%%%%%%%%%-------------------------------------------------- 
Previous test campaigns have demonstrated that reflection gratings operated in an extreme off-plane mount can be measured for soft x-ray diffraction efficiency using a beamline facility suitable for short-wavelength reflectometry \cite{Tutt16,Marlowe16,Miles18,McCoy20}. 
The experiments described here took place at beamline 6.3.2 of the ALS, which provides a highly-coherent beam of radiation tunable over extreme UV and soft x-ray wavelengths that strikes a stage-mounted optic \cite{als632,Underwood96,Gullikson01}. 
At a distance $L \approx 235$~mm away from the point of incidence on the grating, a photodiode detector attached to staging can be used to measure the intensity of propagating orders, which are spaced along the dispersion direction by a distance $\lambda L / d$ as illustrated in Fig.~\ref{fig:conical_reflection}. 
Absolute diffraction efficiency in the $n^{\text{th}}$ propagating order is measured through $\mathcal{E}_n \equiv \mathcal{I}_n / \mathcal{I}_{\text{inc}}$, where $\mathcal{I}_n$ and $\mathcal{I}_{\text{inc}}$ are noise-subtracted intensity measurements of the $n^{\text{th}}$ diffracted beam and the incident beam, respectively, which can be gathered for each order using a vertical, 0.5-mm-wide slit to mask the detector \cite{Miles18,McCoy20}. 
Although this beam is s-polarized to a high degree, x-ray reflection gratings have been demonstrated experimentally to have a polarization-insensitive efficiency response for extreme off-plane geometries \cite{Marlowe16}. 

With the SCIL imprint described in section~\ref{sec:grating_fab} emulating the UV-NIL replica tested by Miles,~et~al.~\cite{Miles18}, diffraction-efficiency testing was carried out in a nearly identical geometry where the half-cone opening angle is $\gamma \approx 1.7^{\circ}$ while the azimuthal incidence angle, $\alpha$, is close to the nominal blaze angle of $\delta = 29.5^{\circ}$ in a near-Littrow configuration. 
The silicon master was tested without a reflective overcoat whereas the inverted SCIL replica was coated with a thin layer of gold to avoid modification of the sol-gel resist by the incident beam, and moreover, to provide a surface with tabulated data for index of refraction and high reflectivity at a $1.7^{\circ}$ grazing-incidence angle. 
This layer was sputter-coated on the replica in an identical fashion to Miles,~et~al.~\cite{Miles18}: 5~nm of chromium was deposited for adhesion followed immediately by 15~nm of gold, without breaking vacuum. 
Because this thickness is several times larger than the $1/\mathrm{e}$ penetration depth in gold at grazing-incidence angles, it is justified to treat this top film as a thick slab in this context \cite{Attwood17,McCoy20}. 

Following the test procedure outlined by Miles,~et~al.~\cite{Miles18}, near-Littrow configurations with $\gamma \approx 1.7^{\circ}$ for both the silicon master and the coated SCIL replica were established at the beamline using principal-axis rotations and in-situ analysis of the diffracted arc. 
The system throw, $L$, was experimentally determined separately for each installed grating by comparing the known detector length to the apparent angular size of the detector as measured by a goniometric scan of the beam at the location of $0^{\text{th}}$ order. 
The arc radius, $r$, was then determined by measuring the locations of propagating orders over a few photon energies and then fitting the data to a half-circle so that $\gamma$ could be inferred from $\sin \left( \gamma \right) = r / L$ \cite{Miles18,McCoy20}. 
Using the $x$-distance between the direct beam and the center of the fitted arc, $\Delta x_{\text{dir}}$, $\alpha$ was measured using $\sin \left( \alpha \right) = \Delta x_{\text{dir}} / r$ before similar calculations described by McCoy,~et~al.~\cite{McCoy20} were carried out to cross-check measured principal-axis angles with $\gamma$ and $\alpha$. 
\begin{table}[t]
 \centering
 \begin{tabular}{cccc}
 \cline{1-3}
 measured parameter               & \ master         		  & \ replica                  \\ \hline
 $L$                              & \ $234.7 \pm 3.0$ mm      & \ $235.6 \pm 3.0$ mm       \\
 $r$                              & \ $6.98 \pm 0.08$ mm      & \ $7.20 \pm 0.14$ mm       \\
 $\Delta x_{\text{dir}}$ 		  & \ $2.80 \pm 0.03$ mm      & \ $3.68 \pm 0.07$ mm       \\
 $\gamma$             			  & \ $1.71 \pm 0.03^{\circ}$ & \ $1.75 \pm 0.04^{\circ}$  \\
 $\alpha$            			  & \ $23.7 \pm 0.7^{\circ}$  & \ $30.7 \pm 0.9^{\circ}$   \\ \hline
 \end{tabular}
 \caption{Measured diffracted arc parameters for the silicon master and the coated SCIL replica in their respective test configurations.}\label{tab:arc_params}
 \end{table}
These measured parameters are listed in Table~\ref{tab:arc_params} for both the silicon master and the coated SCIL replica. 
By the scalar equation for blaze wavelength
\begin{equation}\label{eq:blaze_wavelength}
 \lambda_b = \frac{d \sin \left( \gamma \right)}{n} \left[\sin \left( \alpha \right) + \sin \left( 2 \delta - \alpha \right) \right] \approx \frac{2 d \gamma \sin \left( \delta \right)}{n} \left( 1 - \frac{|\delta - \alpha|^2}{2} \right),
 \end{equation}
where radiation is preferentially diffracted to an angle $\beta = 2 \delta - \alpha$ in Eq.~(\ref{eq:off-plane_incidence_orders}), $\mathcal{E}_n$ for propagating orders with $n=2$ and $n=3$ are expected to maximize in the spectral range 440~eV to 900~eV for a grating with $d \lessapprox 160$~nm in a near-Littrow configuration with $\gamma \approx 1.7^{\circ}$. 
The approximate expression for $\lambda_b$, which is valid for small values of $\gamma$ and $|\delta - \alpha|$, suggests that the locations of peak orders are most sensitive to $\delta$ and $\gamma$ in an extreme off-plane mount rather than $\alpha$ provided that $|\delta - \alpha| \ll 1$~radian, which describes a near-Littrow configuration. 
With both gratings loosely satisfying this condition for $\alpha$, the grating geometries listed in Table~\ref{tab:arc_params} were employed for testing.

Experimental data for $\mathcal{E}_n$ were gathered as a function of photon energy over the range 440~eV to 900~eV in the test configurations summarized in Table~\ref{tab:arc_params}. 
Following Miles,~et~al.~\cite{Miles18}, $\mathcal{I}_n$ for each photon energy was measured using the masked photodiode by scanning the diffracted arc horizontally, in 50~$\mu$m steps, and then determining the maximum of each diffracted order; $\mathcal{I}_{\text{inc}}$ for each photon energy was measured in an analogous way, with the grating moved out of the path of the beam. 
\begin{figure}
 \centering
 \includegraphics[scale=0.26]{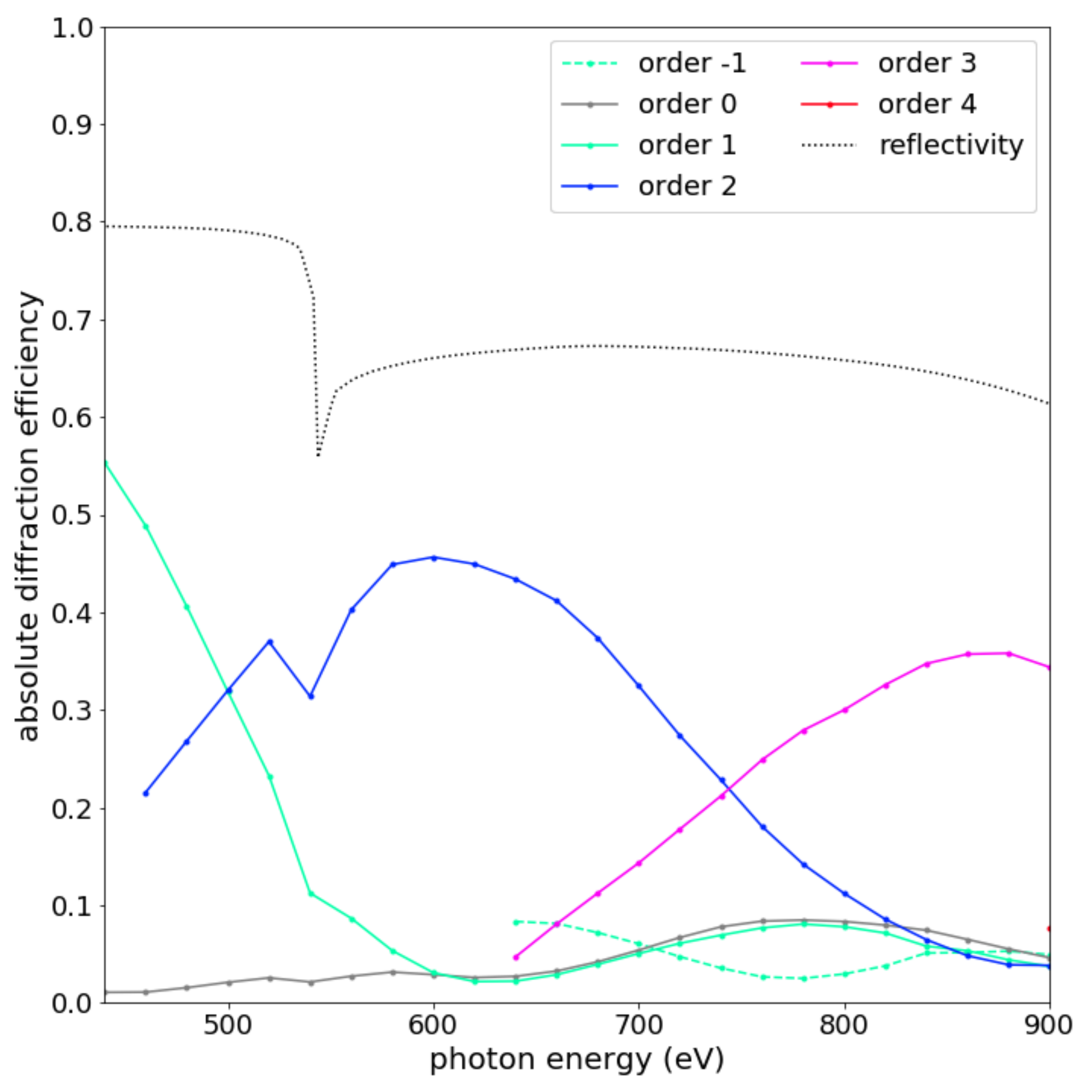}
 \includegraphics[scale=0.26]{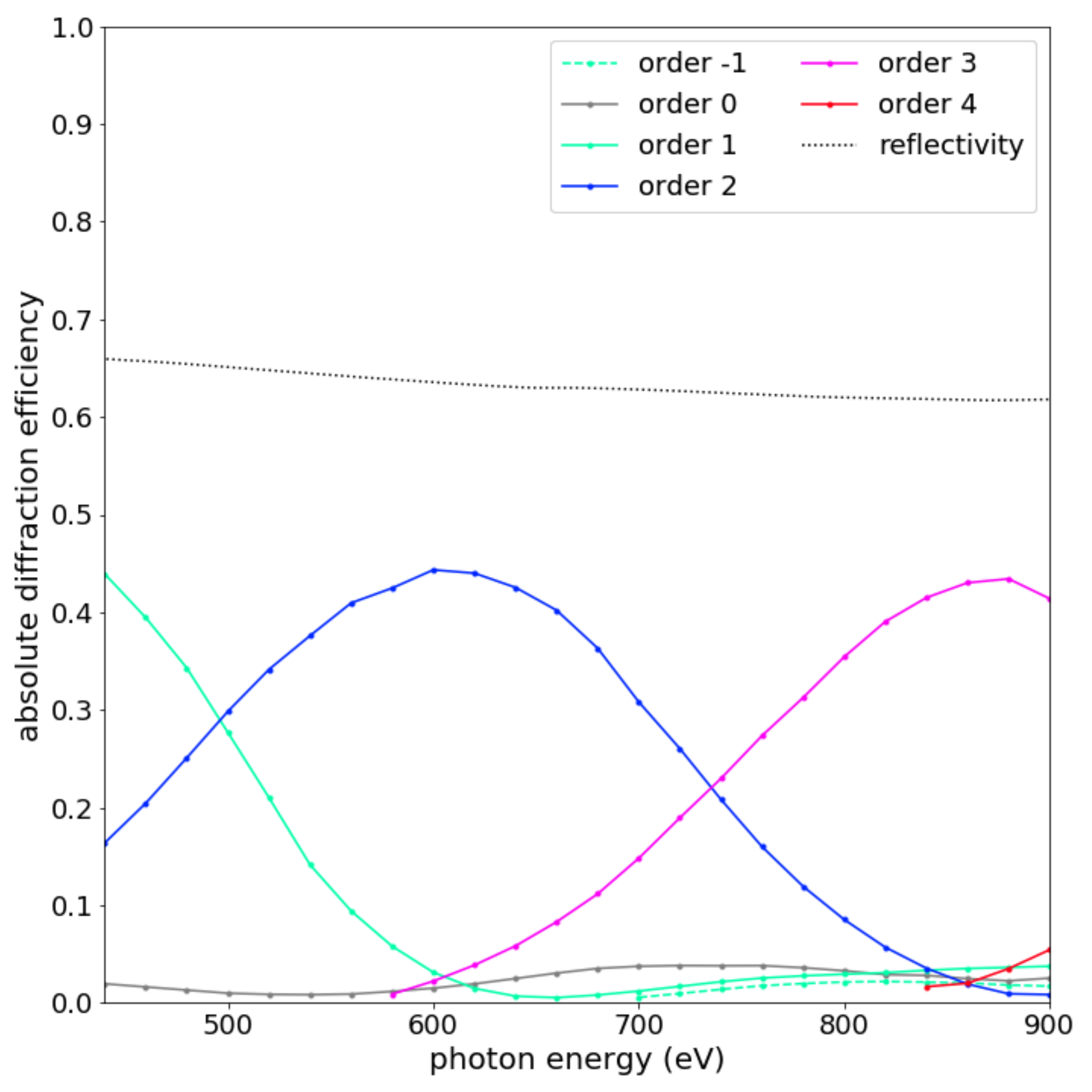}
 \caption{Measured diffraction-efficiency data for the silicon master (\emph{left}) and the gold-coated SCIL replica (\emph{right}) in geometrical configurations described by the parameters listed in Table~\ref{tab:arc_params} compared to Fresnel reflectivity at the facet incidence angle in each case.}\label{fig:abs_eff} 
 \end{figure} 
Through $\mathcal{I}_n / \mathcal{I}_{\text{inc}}$, $\mathcal{E}_n$ was measured every 20~eV between 440~eV and 900~eV for bright propagating orders that are characteristic of each grating's blaze response. 
These results for both the silicon master and the SCIL replica are plotted in Fig.~\ref{fig:abs_eff} and compared to Fresnel reflectivity for silicon with 3~nm of native SiO$_2$ and a thick slab of gold, respectively. 
In an identical fashion to McCoy,~et~al.~\cite{McCoy20}, Fresnel reflectivity was treated using standard-density index of refraction data from the LBNL Center for X-ray Optics on-line database \cite{cxro_data} with a grazing-incidence angle $\zeta$ determined from $\sin \left( \zeta \right) = \sin \left( \gamma \right) \cos \left( \delta - \alpha \right)$, using measured values for $\gamma$, $\alpha$ and $\delta$ (or $\delta'$). 
Peak-order, absolute efficiency ranges from 40-45\% for both gratings or equivalently, 65-70\% measured relative to the reflectivity in each case, which is comparable to the results reported by Miles,~et~al.~\cite{Miles18} for the corresponding UV-NIL replica.

\section{Analysis and Discussion}\label{sec:discussion}
%%%%%%%%%%%%%%%%%%%%%%%%%%%%%%%%%%%%%%%%%--------------------------------------------------
The soft x-ray diffraction-efficiency measurements presented in section~\ref{sec:beamline_testing} demonstrate that both the silicon master and the SCIL replica exhibit a significant blaze response in a near-Littrow, grazing-incidence configuration. 
Using these data, the following analysis seeks to constrain the impact of resist shrinkage on blaze angle in the SCIL replica by comparing measured, single-order efficiency curves to those predicted by theoretical models for diffraction efficiency. 
These models were produced with the aid of the software package \textsc{PCGrate-SX} version 6.1, which solves the Helmholtz equation through the integral method for a custom grating boundary and incidence angles input by the user \cite{pcgrate,Goray10}. 
Based on the findings of Marlowe,~et~al.~\cite{Marlowe16}, which verify that x-ray reflection gratings are polarization-insensitive for extreme off-plane geometries, the incident radiation is treated as a plane wave with transverse-electric polarization relative to the groove direction; the direction of the wave vector, as illustrated in Fig.~\ref{fig:conical_reflection}, is defined by the angles $\gamma$ and $\alpha$ listed in Table~\ref{tab:arc_params}. 
The choice of grating boundary for the silicon master and the SCIL replica follows from the considerations presented in subsections \ref{sec:silicon_master} and \ref{sec:scil_replica}, respectively, along with AFM measurements of the tested gratings. 
In each case, the grating boundary is taken to be perfectly conducting in \textsc{PCGrate-SX} while the overall response is modulated by Fresnel reflectivity to yield a predicted result for absolute diffraction efficiency. 
Considering that the $\lessapprox$0.5-mm cross-sectional diameter of the beam projects to tens of millimeters at grazing incidence, and that the point of incidence is the central grooved region of each grating, the groove spacing in each case is taken to be $d=159.125$~nm, which is the nominal average of the variable-line-space profile described in section~\ref{sec:grating_fab}. 

\subsection{Silicon Master}\label{sec:silicon_master}
%%%%%%%%%%%%%%%%%%%%%%%%%%%%%%%%%%%%%%%%%--------------------------------------------------
As a point of reference for examining resist shrinkage in the SCIL replica, the diffraction-efficiency results for the silicon master from the left panel of Fig.~\ref{fig:abs_eff} are compared to various \textsc{PCGrate-SX} models that are based on the wet-etched grating topography described in section~\ref{sec:grating_fab}. 
\begin{figure}
 \centering
 \includegraphics[scale=0.2]{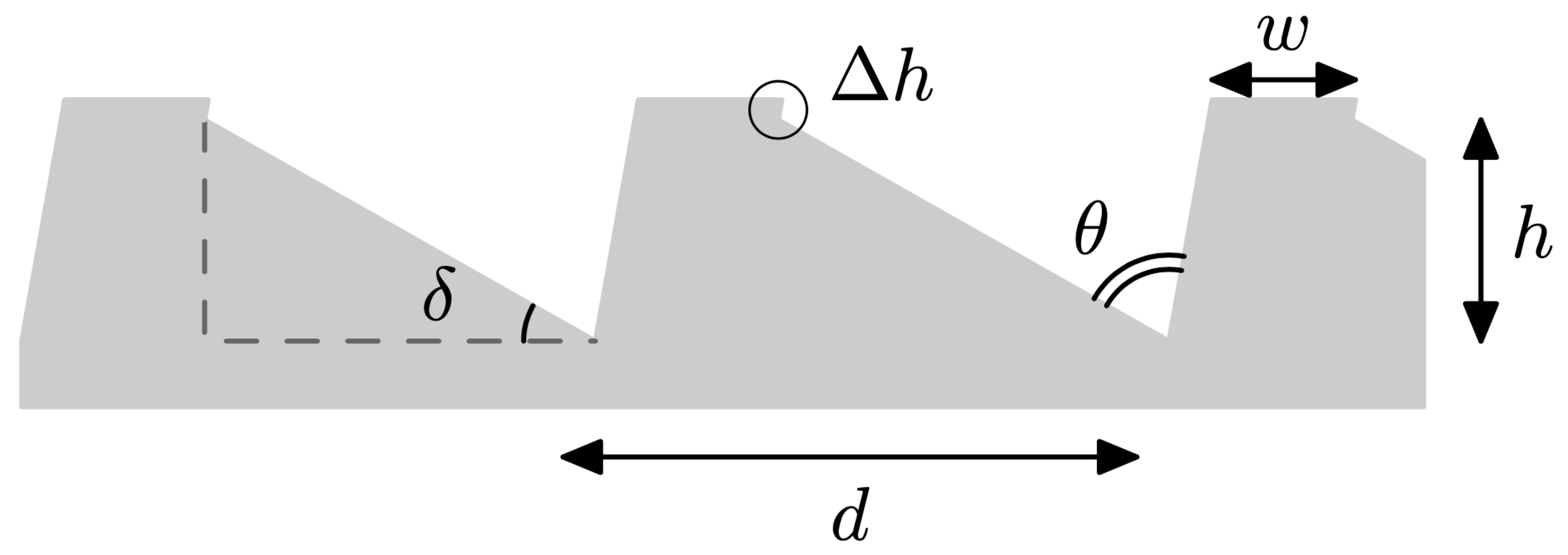} 
 \caption{Schematic illustration of the silicon master cross-section with $\delta = 29.5^{\circ}$ as the blaze angle and $\theta \approx 70.5^{\circ}$ defined by the crystal structure of silicon. At a groove spacing of $d \lessapprox 160$~nm, the flat-top regions have widths $w \gtrapprox 30$~nm as a result of the etch undercut while the groove depth is $h \approx 65-70$~nm by Eq.~(\ref{eq:groove_depth}). Indicated by the circle, the indented portion of the etched topography cannot be described with a functional form for diffraction-efficiency analysis.}\label{fig:master_profile} 
 \end{figure}
Illustrated in Fig.~\ref{fig:master_profile} and shown under FESEM in Fig.~\ref{fig:master_SEM}, the cross-sectional shape of the grating profile resembles a series of acute trapezoids with flat tops of width $w$ that each protrude a distance $\Delta h$ of a few nanometers so that the groove depth, $h$, is given approximately by 
\begin{equation}\label{eq:groove_depth}
 h \approx \frac{d - w}{\cot \left( \delta \right) - \cot \left( \theta + \delta \right)} + \Delta h 
 \end{equation}
with $\theta \approx 70.5^{\circ}$ defined by the intersection of exposed $\{ 111 \}$ planes and $\delta$ as the active blaze angle. 
Although the depth of these sharp grating grooves could not be verified by AFM due to the moderate aspect ratio of the scanning-probe tip, it is estimated that this quantity falls in the range $h \approx 65-70$~nm based on the expected value of $\delta = 29.5^\circ$ for a $\langle 311 \rangle$-oriented silicon surface. 
Under AFM, facet surface roughness, $\sigma$, measures $\lessapprox 0.4$~nm RMS while the average of 30 blaze angle measurements over a 0.5~$\mu$m by 1~$\mu$m area yields $\delta = 30.0 \pm 0.8^{\circ}$, where the uncertainty is one standard deviation. 
Although these AFM data were gathered with vertical measurements calibrated to a 180-nm standard at the PSU Materials Characterization Laboratory, this blaze angle measurement is limited in its accuracy due to a relatively poor lateral resolution on the order of a few nanometers. 
The measurement is, however, consistent with the nominal value of $\delta = 29.5^\circ$ and is considered a reasonable estimation for the blaze angle of the silicon master. 

\begin{figure*}
 \centering
 \includegraphics[scale=0.74]{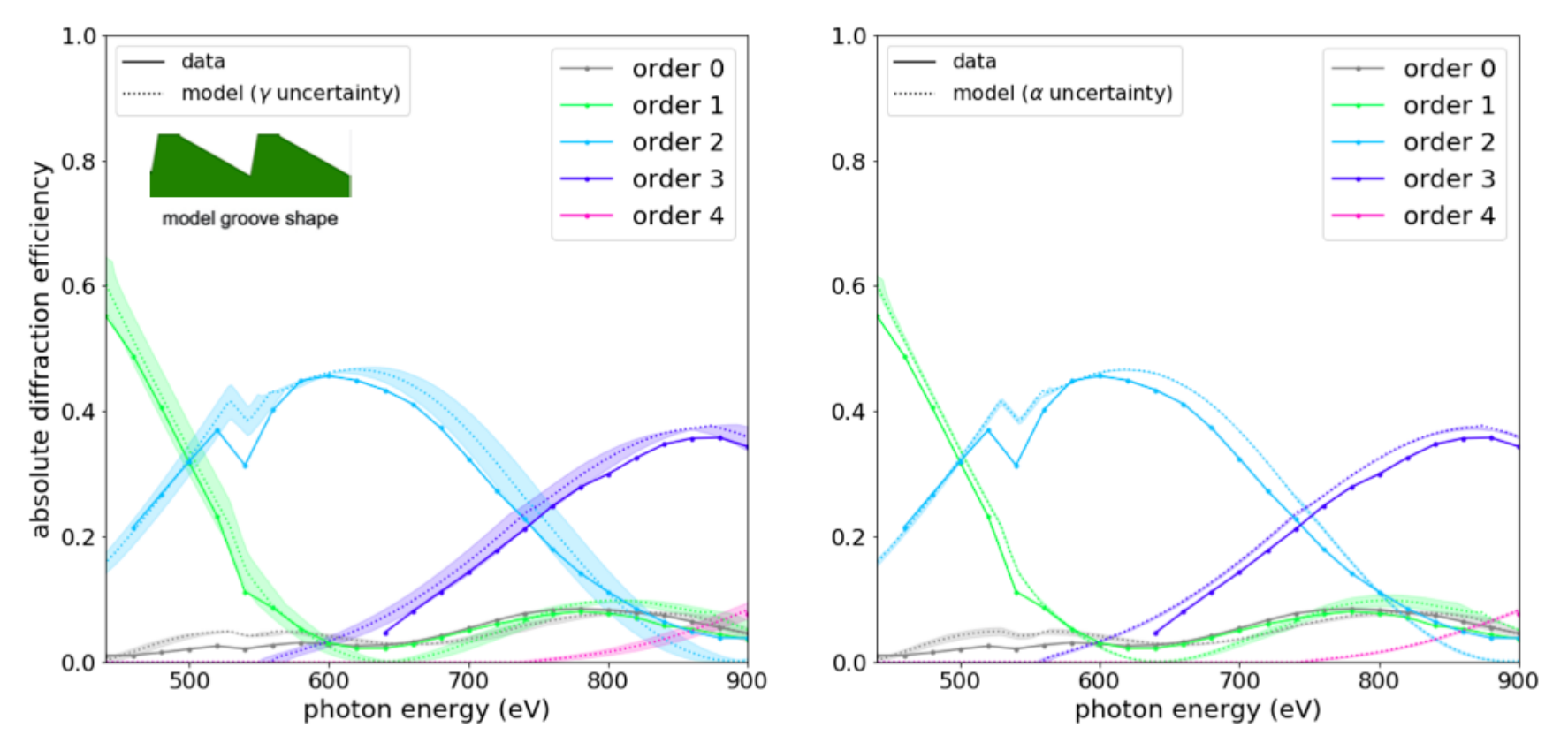}
 \caption{Measured diffraction-efficiency data for the silicon master from the left panel of Fig.~\ref{fig:abs_eff} compared to \textsc{PCGrate-SX} models that assume a groove profile similar to the wet-etched topography described in section~\ref{sec:grating_fab}, with sawtooth angles $\delta=29.5^{\circ}$ and $180^{\circ} - \theta - \delta \approx 80^{\circ}$, a flat-top width of $w = 35$~nm, a nub-protrusion height of $\Delta h = 3$~nm and an overall grove depth of $h \approx 67$~nm by Eq.~(\ref{eq:groove_depth}). 
 In the left and right panels, respectively, $\gamma$ and $\alpha$ are allowed to vary at levels of $\pm 0.03^{\circ}$ and $\pm 0.7^{\circ}$, which are represented by shaded uncertainty swaths.}\label{fig:model_si1}
 \end{figure*}

From the above considerations, the grating boundary used for \textsc{PCGrate-SX} modeling was defined using the trapezoid-like groove shape shown in the inset of Fig.~\ref{fig:model_si1}, with nominal sawtooth angles of $\delta = 29.5^{\circ}$ and $80^{\circ}$, a flat-top width of $w = 35$~nm, a nub-protrusion height of $\Delta h = 3$~nm and a groove depth of $h \approx 67$~nm that follows from Eq.~(\ref{eq:groove_depth}). 
In both panels of Fig.~\ref{fig:model_si1}, the model that utilizes the nominal values $\gamma = 1.71^{\circ}$ and $\alpha = 23.7^{\circ}$ is plotted using dotted lines for each diffracted order shown, with uncertainties listed in Table~\ref{tab:arc_params} represented as shaded swaths.  
\begin{figure*}
 \centering
 \includegraphics[scale=0.74]{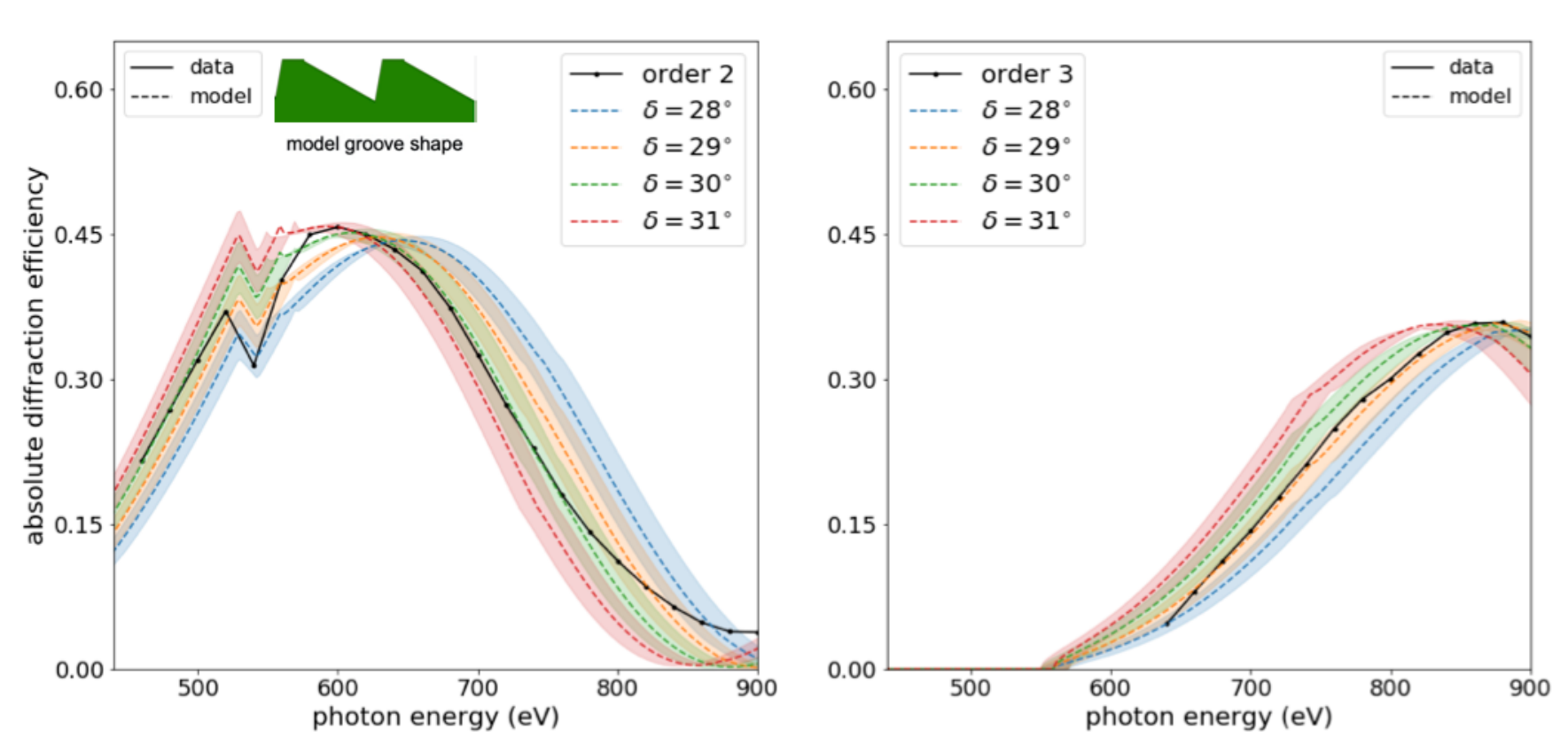}
 \caption{Measured diffraction-efficiency data in orders $n=2$ and $n=3$ for the silicon master compared to \textsc{PCGrate-SX} models with $28^{\circ} \leq \delta \leq 31^{\circ}$ that are normalized to match the data in terms of peak efficiency while the shaded swaths represent the $\pm 0.03^{\circ}$ uncertainty in $\gamma$. These results indicate that $\delta$ for the silicon master is close to the nominal value of $\delta = 29.5^{\circ}$.}\label{fig:model_si2} 
 \end{figure*}
These results show that the constrained geometry leads to the production of models that roughly match the experimental data. 
Mismatches between the models and the data may be in part due to the detailed shape of nubs atop of each groove, which cannot be described with a functional form as illustrated in Fig.~\ref{fig:master_profile}. 
Although this limits the accuracy of the \textsc{PCGrate-SX} models utilized, the model uncertainty swaths indicate that $\gamma$ serves to shift the centroids of peak orders (i.e. the photon energy equivalent to $\lambda_b$) while $\alpha$ has a small impact as expected from Eq.~(\ref{eq:blaze_wavelength}). 

With the centroids of the efficiency curves shown in Fig.~\ref{fig:model_si1} depending directly on the blaze angle by Eq.~(\ref{eq:blaze_wavelength}), a series of models with $28^{\circ} \leq \delta \leq 31^{\circ}$ in steps of $1^{\circ}$ are compared to $n=2$ and $n=3$ absolute-efficiency data in Fig.~\ref{fig:model_si2}. 
In each of these models, $w = 35$~nm and $\Delta h = 3$~nm are fixed while the sawtooth angles vary as $\delta$ and $180^{\circ} - \theta - \delta$ with the overall groove depth, $h$, following from Eq.~(\ref{eq:groove_depth}). 
The modeled efficiency in each case, which assumes a perfectly smooth grating boundary due to the small RMS facet roughness measured by AFM, was normalized to match the peak efficiency of the measured data so that the peak-centroid positions could be compared. 
Dotted lines represent the nominal model with $\gamma = 1.71^{\circ}$ and $\alpha = 23.7^{\circ}$ while the shaded swaths show the $\pm 0.03^{\circ}$ uncertainty in $\gamma$. 
These results support the expectation that the blaze angle of the silicon master is in the neighborhood of the nominal value of $\delta = 29.5^{\circ}$ as well as the AFM-measured value of $\delta = 30.0 \pm 0.8^{\circ}$. 

\subsection{SCIL Replica}\label{sec:scil_replica}
%%%%%%%%%%%%%%%%%%%%%%%%%%%%%%%%%%%%%%%%%--------------------------------------------------
In a similar manner to Fig.~\ref{fig:model_si2} for the silicon master, the experimental data from the right panel of Fig.~\ref{fig:abs_eff} are compared to several \textsc{PCGrate-SX} models with varying blaze angle, $\delta'$, in order to evaluate resist shrinkage in the SCIL replica. 
\begin{figure*}
 \centering
 \includegraphics[scale=0.44]{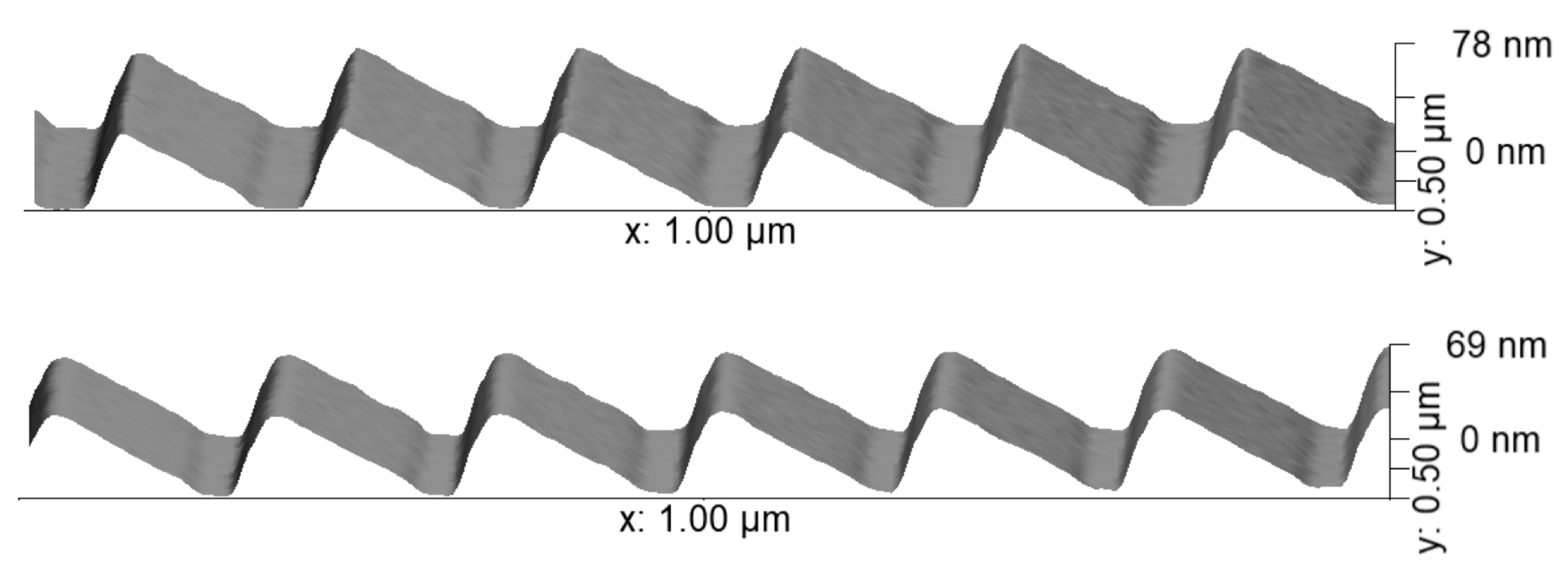}
 \caption{AFM images of a grating imprint with a groove spacing of $d \lessapprox 160$~nm in sol-gel resist, as in Fig.~\ref{fig:SCIL_FESEM}. The bare imprint (\emph{top}) has facet roughness and average blaze angle measuring $\sigma \approx 0.6$~nm RMS and $\delta' = 27.9 \pm 0.7^{\circ}$, respectively. The sputter-coated imprint (\emph{bottom}) yields $\sigma \approx 0.8$~nm RMS while the average blaze angle is statistically consistent with $\delta' = 28.4 \pm 0.8^{\circ}$.}\label{fig:SCIL_AFM} 
 \end{figure*}
Such a grating imprint in sol-resist produced using the methodology described in section~\ref{sec:grating_fab} is shown under AFM in the top panel of Fig.~\ref{fig:SCIL_AFM} while an identical grating following the sputtering deposition described in section~\ref{sec:beamline_testing} is shown in the bottom panel. 
The average blaze angle from 30 measurements over these 0.5~$\mu$m by 1~$\mu$m areas measures $\delta' = 27.9 \pm 0.7^{\circ}$ for the bare imprint and $\delta ' = 28.4 \pm 0.8^{\circ}$ following the coating. 
These measurements, which are consistent with one another to one standard deviation, give $\delta' / \delta = 0.93 \pm 0.03$ and $\delta' / \delta = 0.95 \pm 0.04$ as a reduction in blaze angle relative to $\delta = 30.0 \pm 0.8^{\circ}$ measured for the silicon master. 
The statistical consistency between these two measurements suggests that coating effects had a minimal impact on the blaze angle and that $\delta' / \delta$ constrained from diffraction-efficiency testing results is expected to be indicative of resist shrinkage alone. 

Unlike the silicon master profile illustrated in Fig.~\ref{fig:master_profile}, the inverted topography of the SCIL replica features a relatively sharp apex and a flat-bottom portion of width $w$, which is largely shadowed in a near-Littrow configuration. 
With \textsc{PCGrate-SX} simulations showing that only the active blaze angle significantly affects the results in terms of peak-order centroids in such a geometry, the groove profile for diffraction-efficiency modeling is treated as an ideal sawtooth with a sharp, $90^{\circ}$ apex angle and no flat-bottom portion, which yields a groove depth of $h \approx 66$~nm. 
As in Fig.~\ref{fig:model_si2} for the silicon master, these models assume perfectly smooth surfaces and are normalized to the data in terms of peak efficiency in order to compare peak centroids. 
\begin{figure*}
 \centering
 \includegraphics[scale=0.74]{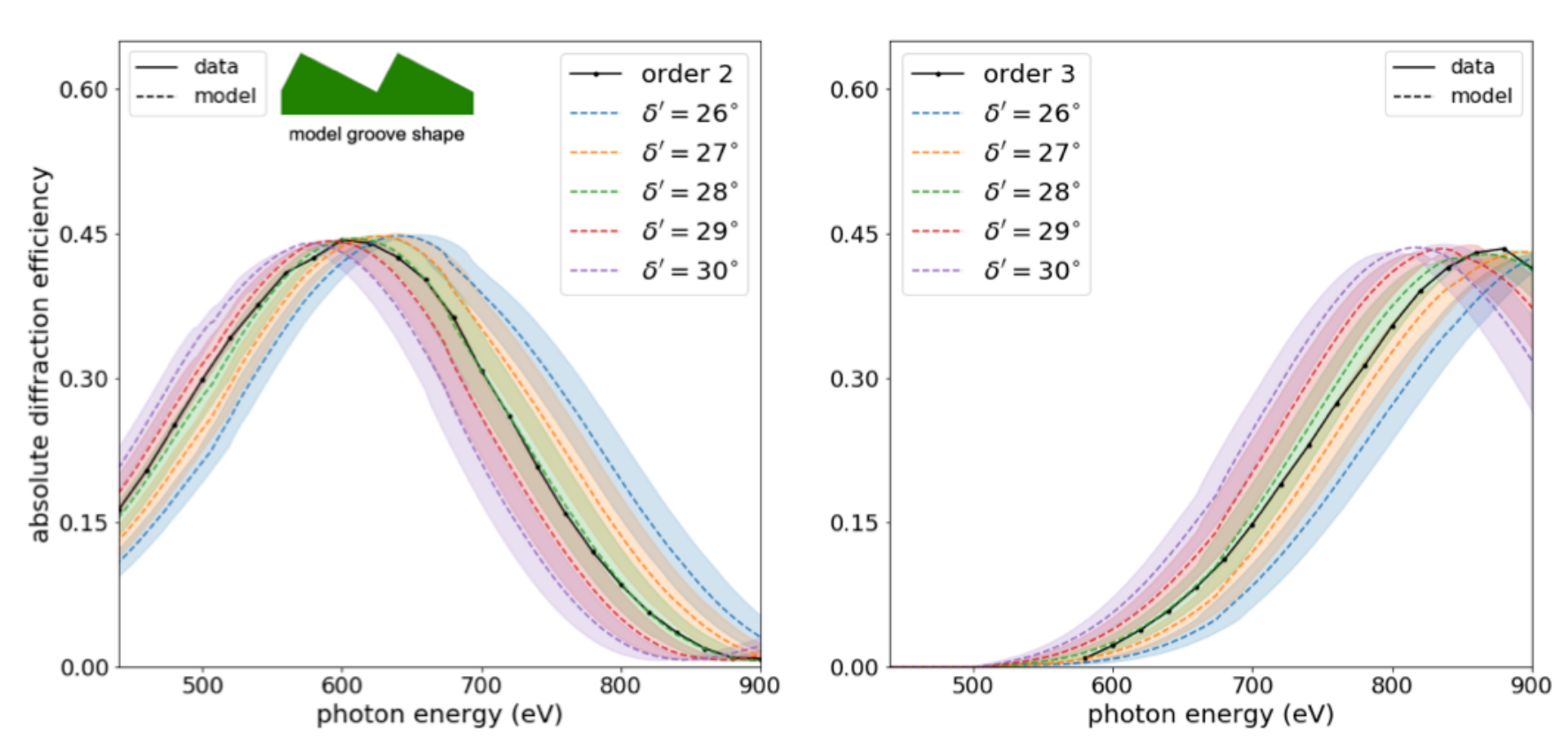}
 \caption{Measured diffraction-efficiency data in orders $n=2$ and $n=3$ for the coated SCIL replica compared to \textsc{PCGrate-SX} models that assume an ideal sawtooth with blaze angles ranging between $26^{\circ} \leq \delta' \leq 30^{\circ}$, which have been normalized to match the data. These results show that the measured data most closely match a grating with $\delta' = 28^{\circ}$.}\label{fig:model_scil2} 
 \end{figure*} 
The outcome is presented in Fig.~\ref{fig:model_scil2} where the diffraction-efficiency data for the SCIL replica in orders $n=2$ and $n=3$ are each plotted against five \textsc{PCGrate-SX} models with $26^{\circ} \leq \delta' \leq 30^{\circ}$ in steps of $1^{\circ}$, all with $\alpha = 30.7^{\circ}$ and $\gamma = 1.75 \pm 0.04^{\circ}$ from Table~\ref{tab:arc_params}, with the latter represented by uncertainty swaths. 
It is apparent from Fig.~\ref{fig:model_scil2} that the data are most consistent with the $\delta' = 28^{\circ}$ model, as expected from AFM measurements. 
In order to interpret this result in the context of SCIL processing, $\delta' \approx 28^{\circ}$ is compared to an approximate model for resist shrinkage that is considered in the following discussion. 

To formulate a simple model resist shrinkage, it is first assumed that shrinkage effects in the SCIL stamp can be neglected, which is expected due to the high intrinsic cross-link density of X-PDMS \cite{Verschuuren19}. 
The profile of the imprinted blazed grating, without resist shrinkage, is considered to be composed of a series of groove facets with spacing $d \lessapprox 160$~nm that resembles the inverse of the silicon master described in section~\ref{sec:grating_fab}. 
These facets are separated from one another by the distance $w \gtrapprox 30$~nm defined in Fig.~\ref{fig:master_profile} so that the base of each groove facet has a width $b \approx d - w \lessapprox 130$~nm, which is assumed to be a small enough size scale for material relaxation in sol-gel resist. 
\begin{figure}
 \centering
 \includegraphics[scale=0.28]{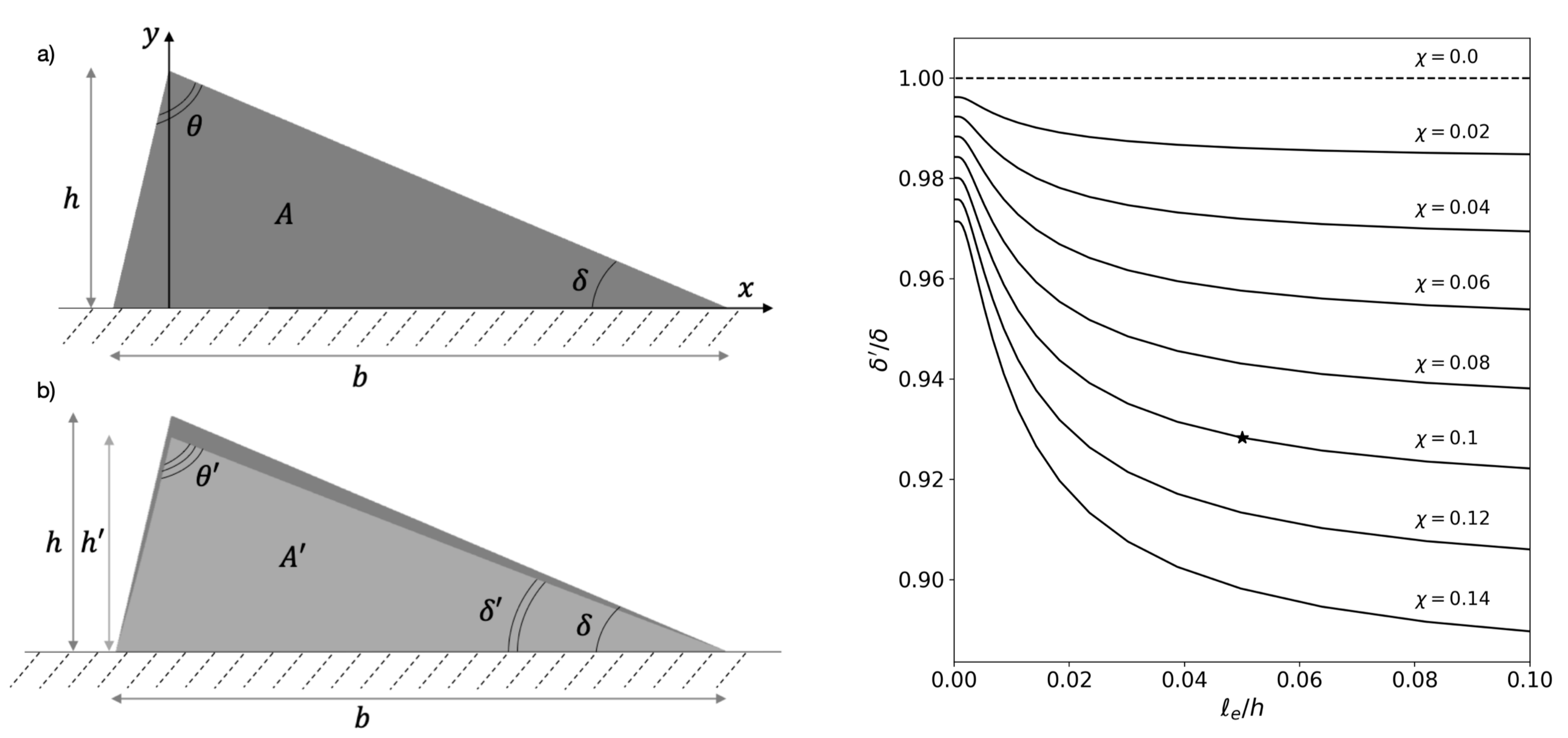}
 \caption{Approximate model for resist shrinkage with $y=0$ representing a fixed boundary defined by the residual layer. \emph{Left:} a) The original facet shape has a blaze angle $\delta = 29.5^{\circ}$, an apex angle $\theta \approx 70.5^{\circ}$ and an area $A$. b) A shrunken facet is generated by dividing the original facet shape into 1000 layers along the $y$-direction and then requiring that the area of each is reduced to $A'= 0.9 A$ with $\chi = 0.1$ while the ratio between lateral and vertical shrinkage varies with $y$ according to Eq.~(\ref{eq:exp_shrink}) for $\ell_e / h = 0.05$. \emph{Right:} Reduced blaze angle predicted by model relative to the initial blaze angle, $\delta' / \delta$, as a function of $\ell_e / h$ for various values of $\chi$. The marked star indicates $\chi = 0.1$ and $\ell_e / h = 0.05$ used for the illustrated model.} \label{fig:shrinkage} 
 \end{figure}
As illustrated in in Fig.~\ref{fig:shrinkage}(a), the shallow side of the facet is assigned the nominal value of $\delta = 29.5^{\circ}$ while the effect of the protruding nubs on the silicon master is ignored for simplicity so that the groove depth with $\Delta h = 0$ is $h \lessapprox 67$~nm by Eq.~(\ref{eq:groove_depth}). 
Simulations of resist shrinkage in UV-NIL based on continuum mechanics of elastic media indicate that on average, a volume element $V$ shrinks to $V' = V \left( 1 - \chi \right)$ with $\chi$ as the fractional loss in volume \cite{Shibata10,Horiba_2012}. 
In this regard, the residual layer of resist that exists beneath the groove facets is expected to experience reduction in thickness alone.  Stress-induced substrate deformation from this laterally-constrained shrinkage is considered to be negligible owing to the 1-mm thickness of the silicon wafer used for the grating replica.

The residual layer effectively serves as a fixed boundary for the shrinking groove facets, which retain their original groove spacing, $d$, throughout the process of resist shrinkage \cite{Horiba_2012}. 
As such, shrinkage in each of these groove facets is assumed to manifest as a reduction in cross-sectional area due to the inability of the material network to relax over large groove lengths.  
Without knowledge of the elastic properties of sol-gel resist or the details of its thermodynamical shrinkage mechanism, the simple resist-shrinkage model presented here stems from the assumption that throughout each imprinted groove facet, the reduction in cross-sectional area from $A = b h / 2$ to $A' = A \left( 1 - \chi \right)$ is uniform in magnitude while the ratio of lateral shrinkage to vertical shrinkage, $S$, varies spatially according to 
\begin{equation}\label{eq:exp_shrink}
 S = 1 - \mathrm{e}^{- y / \ell_e} \quad \text{for} \ \ 0 \leq y \leq h  
 \end{equation}
with $\ell_e$ as an arbitrary $1 / \mathrm{e}$ length scale for $S$ approaching unity as $y$ increases toward $h$. 
By introducing $s_x = 1 - S f$ and $s_y = 1 - f$ as functions of position that describe shrinkage in the $x$ and $y$ directions shown in Fig.~\ref{fig:shrinkage}(a) and then requiring $s_x s_y = 1 - \chi$, it is found that  
\begin{equation}\label{eq:quad_solution}
 f = \frac{1 + S - \sqrt{(1+S)^2 - 4 S \chi}}{2 S} \quad \text{for} \ \ 0 \leq S \leq 1
 \end{equation}
parameterizes $s_x$ and $s_y$. 
These expressions are incorporated into the resist-shrinkage model by first considering the original groove facet shape shown in Fig.~\ref{fig:shrinkage}(a) to be composed of 1000 rectangular layers, each with an identical, thin, vertical thickness. 
A shrunken facet profile is produced by requiring the area of each of these layers to be reduced according to $s_x$ and $s_y$ for specified values of $\chi$ and $\ell_e$. 

Figure~\ref{fig:shrinkage}(b) shows a shrunken facet profile predicted for $\chi = 0.1$ and $\ell_e = 0.05 h$ where the blaze angle is reduced to $\delta' \approx 0.93 \delta $ while the groove depth shrinks to $h' \approx 0.91 h$ as the apex angle widens with $\theta' \approx 1.05 \theta$. 
Because the facet features curvature near its base and flattens to a linear slope as $y$ becomes larger than $\ell_e$, $\delta'$ is measured from the upper half of the facet, where $S \lessapprox 1$ for relatively small values of $\ell_e / h$. 
The quantity $\delta' / \delta$ determined in this way is plotted as a function of $\ell_e / h$ for various values of $\chi$ in the right panel of Fig.~\ref{fig:shrinkage}, where the marked star indicates $\chi = 0.1$ and $\ell_e / h = 0.05$ for the illustrated model. 
Despite $\ell_e / h$ remaining poorly constrained without measurements for $h'/h$ and $\theta' / \theta$, the comparison between the resist-shrinkage model just presented and $\delta' / \delta \approx 0.93$ determined from diffraction-efficiency analysis along with AFM measurements supports the hypothesis stated in section~\ref{sec:intro} that the level of volumetric shrinkage for a 90$^{\circ}$C-treated sol-gel imprint is approximately 10\%. 
Although this analysis does not tightly constrain $\delta'$, it does demonstrate that the SCIL replica functions as a blazed grating with a facet angle reduced by $\sim$2$^{\circ}$ relative to the silicon master, which has been shown to exhibit a blaze angle of $\delta \approx 30^{\circ}$, giving a value for $\delta' / \delta$ that is consistent with a typical shrunken facet with $\chi \approx 0.1$. 

\section{Summary and Conclusions}\label{sec:conclusion}
%%%%%%%%%%%%%%%%%%%%%%%%%%%%%%%%%%%%%%%%%--------------------------------------------------
This paper describes a SCIL process for patterning blazed grating surface-relief molds in \textsc{NanoGlass T1100}, a thermodynamically-curable, silica sol-gel resist, and characterizes the impact of resist shrinkage induced by a $90^{\circ}$C post-imprint treatment through diffraction-efficiency testing in the soft x-ray supported by AFM measurements of the blaze angle. 
An imprinted grating that features the inverse topography of the wet-etched silicon master template was sputter-coated with gold, using chromium as an adhesion layer, before being tested for diffraction efficiency in an extreme off-plane mount at beamline 6.3.2 of the ALS. 
By testing the silicon master in a similar configuration and comparing the results of both gratings to theoretical models for diffraction efficiency, it was found that the response of the coated SCIL replica is consistent with a reduced blaze angle of $\delta' \approx 28^{\circ}$ whereas the silicon master yields diffraction-efficiency results characteristic of a nominal $\langle 311 \rangle$ blaze angle with $\delta \approx 30^{\circ}$. 
According to an approximate model formulated for resist shrinkage, this outcome supports the hypothesis that the replicated grating experienced volumetric shrinkage in the sol-gel resist on the level of 10\%. 
The result serves as experimental evidence for sol-gel resist shrinkage impacting the performance of an x-ray reflection grating in terms of its ability to maximize diffraction efficiency for a specific diffracted angle. 
Monitoring this effect is particularly relevant for instrument development in astrophysical x-ray spectroscopy that relies on the production of large numbers of identical gratings, where resist shrinkage should be compensated for in the master grating to ensure that the replicas perform as expected \cite{Miles19b,Tutt18,McEntaffer19}. 
Although the \textsc{AutoSCIL} production platform provides an avenue for high-volume production of grating imprints, sputter-coating is limited in its throughput, and moreover, the impact of ion bombardment on the sol-gel network has not been investigated. 
This motivates the pursuit of alternative deposition processes that are both capable of high throughput and compatible with sol-gel resist.  

\section*{Funding}
%%%%%%%%%%%%%%%%%%%%%%%%%%%%%%%%%%%%%%%%%--------------------------------------------------
National Aeronautics and Space Administration (NNX16AP92H, 80NSSC17K0183); U.S. Department of Energy (DE-AC02-05CH11231).

%OSA participates in \href{https://www.crossref.org/fundingdata/}{Crossref's Funding Data}, a service that provides a standard way to report funding sources for published scholarly research. To ensure consistency, please enter any funding agencies and contract numbers from the Funding section in Prism during submission or revisions. If exact wording for a funder is required, this may be added to the Acknowledgment section, even if it duplicates information in the funding information. 

\section*{Acknowledgments}
This research was supported by NASA Space Technology Research Fellowships and used resources of the Nanofabrication Laboratory and the Materials Characterization Laboratory at the Penn State Materials Research Institute, Philips SCIL Nanoimprint Solutions and beamline 6.3.2 of the Advanced Light Source, which is a DOE Office of Science User Facility under contract no.\ DE-AC02-05CH11231.

\section*{Disclosures}
%%%%%%%%%%%%%%%%%%%%%%%%%%%%%%%%%%%%%%%%%--------------------------------------------------
The authors declare no conflicts of interest.

%%%%%%%%%%%%%%%%%%%%%%% References %%%%%%%%%%%%%%%%%%%%%%%%%

%%%%%%%%%% If using BibTeX:
\bibliography{sample}

%%%%%%%%%% If preparing manually:
% \newcommand{\enquote}[1]{``#1''}

% \bibitem{Zhang:14}
% Y.~Zhang, S.~Qiao, L.~Sun, Q.~W. Shi, W.~Huang, L.~Li, and Z.~Yang,
%   \enquote{Photoinduced active terahertz metamaterials with nanostructured
%   vanadium dioxide film deposited by sol-gel method,}
%   {\protect\JournalTitle{Optics Express}} \textbf{22}, 11070--11078 (2014).

% {Optical Society}, \enquote{{OSA Publishing},}
%   \url{http://www.osapublishing.org}.

% \bibitem{FORSTER2007}
% P.~Forster, V.~Ramaswamy, P.~Artaxo, T.~Bernsten, R.~Betts, D.~Fahey,
%   J.~Haywood, J.~Lean, D.~Lowe, G.~Myhre, J.~Nganga, R.~Prinn, G.~Raga,
%   M.~Schulz, and R.~V. Dorland, \enquote{Changes in atmospheric consituents and
%   in radiative forcing,} in \enquote{Climate Change 2007: The Physical Science
%   Basis. Contribution of Working Group 1 to the Fourth assesment report of
%   Intergovernmental Panel on Climate Change,}  S.~Solomon, D.~Qin, M.~Manning,
%   Z.~Chen, M.~Marquis, K.~B. Averyt, M.~Tignor, and H.~L. Miler, eds.
%   (Cambridge University Press, 2007).

% \end{thebibliography}

\end{document}